\documentstyle[prb,aps,multicol]{revtex} 
\input epsf

\renewcommand{\Re}{{\rm Re}} 
\renewcommand{\Im}{{\rm Im}} 
\begin{document} 
\draft 
\title{Spin-orbit induced anisotropy in the magnetoconductance 
of two-dimensional metals}  
\author{R. Raimondi, M. Leadbeater 
} 
\address{  
INFM e Dipartimento di Fisica 
"E. Amaldi", 
 Universit\`a di Roma Tre, 
	Via della Vasca Navale 84, 00146 Roma, Italy 
} 
\author{P. Schwab}
\address{Institut f\"ur Physik, Univerit\"at Augsburg, 86135 Augsburg Germany }
\author{E. Caroti, C. Castellani}
\address{INFM e Dipartimento di Fisica, 
 Universit\`a ``La Sapienza'',  
	 piazzale A. Moro 2, 00185  Roma  Italy }
	 
\date{\today} 
\maketitle 
\begin{abstract} 
It is shown that the spin-orbit coupling due to structure inversion
asymmetry leads
to a characteristic anisotropy in the magnetoconductance
of two-dimensional metals. Relevance for recent experiments
is discussed.
\end{abstract} 
\pacs{PACS numbers: 72.10-d, 72.15.Rn}

\begin{multicols}{2} 
Over the last five years there has been a significant amount of 
experimental evidence  of a metallic-like temperature dependent
resistivity for a certain electron density range
in Si-MOSFETs\cite{kravchenko95} 
and semiconductor heterostructures. This has called for a better 
understanding of the possible ground states
of two-dimensional (2D) systems. For such an endeavour it is
important from the outset to identify the relevant physical mechanisms
at play. 
A recent summary of the main experimental facts and suggested theoretical 
models can be found in Ref.\cite{abrahams00}.
In the 2D systems which are
experimentally investigated, the spin-orbit interaction, as due to the
lack of structural inversion symmetry\cite{rashba84}, has been proposed
as a potentially relevant mechanism for the observed resistivity behaviour
\cite{pudalov97}. However, there is no consensus at present about
the importance of this spin-orbit interaction.

Magneto-transport experiments are  
very powerful tools in selecting processes which are affected by the 
presence of a magnetic field. In the case of 2D systems one of the most
puzzling features is the magnetoresistance in a parallel magnetic field.
For carrier densities of interest
the resistance  increases by roughly one order of magnitude on the scale
of several Tesla before saturating at high magnetic
fields\cite{simonian97}.
It has been generally assumed that a parallel magnetic field affects the
system only via the coupling to the electron spin, and there is 
experimental evidence supporting this view. 
However the importance of the coupling of the
magnetic field to the orbital motion of the electrons as a consequence of the
finite layer thickness of the 2D structure has also been
proposed\cite{dassarma99}. 

In order to shed light on the problem there has recently been
the proposal that, in the presence of the spin-orbit interaction,  
an in-plane magnetic field gives rise to an anisotropy 
in the magnetoresistance, depending on whether one measures the current in the
direction perpendicular or parallel to the magnetic field\cite{chen00}.
This has stimulated more experimental work, where anisotropy in the
magnetoresistance has been found in $GaAs$ hole systems\cite{papadakis99},
 Si-MOSFETs\cite{pudalov00}, and $GaAs$ electron systems\cite{khrapai00}.
The prediction of Ref.\cite{chen00} has been made in the deeply insulating
limit and this makes a direct comparison with the
experiments a little difficult. 
This fact has motivated us to investigate the interplay
of the spin-orbit interaction and an in-plane magnetic field in the metallic
regime. 
Here we calculate the conductivity in the framework of the Drude-Boltzmann
theory.
We do not worry about the microscopic orign of the scattering which is,
for example, responsible for the strong magnetic field dependence of the 
conductivity.
Our concern is the anisotropy of the conductance and
its behaviour
as a function of the various physical parameters. 
The most important 
findings are: a) the
magnetoresistance is anisotropic; b) the sign of the anisotropy
depends on the type of scattering potential seen by the charge carriers;
c) the anisotropy factor has a maximum as function of the
magnetic field, with the maximum position scaling with the density.
In the following we present details of our calculations and discuss its
possible experimental relevance.

We start from the model Hamiltonian 
\begin{equation}
\label{1}
H= \frac{p^2}{2m}+\alpha {\mathbf {\sigma}}\cdot {\bf p}\wedge{\bf e}_z  -
\frac{1}{2}g\mu_B{\mathbf \sigma}\cdot {\bf B}
\end{equation}
 with $\alpha$ a parameter
describing the spin-orbit interaction due to the confinement field and 
${\bf {\sigma}}$ being a vector whose components are the Pauli matrices.
The unit vector ${\bf e}_z$ is perpendicular to the 2D plane and defines
the $z$-axis. The magnetic field is chosen to lie in $x$-direction.
One finds then two bands with the dispersion
\begin{equation}
E_{\pm}({\bf p}) = p^2/2m \pm \Omega({\bf p}), 
\end{equation}
with $\Omega({\bf p }) = |\alpha {\bf p} \wedge {\bf e}_z -  \omega_s {\bf e}_x | $. 

In the problem we have several energy scales: the Fermi energy $\epsilon_F$
(in the absence of magnetic field with no spin-orbit coupling),
the Zeeman energy $\omega_s= \frac{1}{2} g \mu_B B$ (note that this differs by
a factor two from the standard definition), 
and the spin-orbit energy $\alpha p_F$.
It is useful to make a few estimates. In Si inversion layers one
has a density-of-states 
$N_0=1.59\times 10^{11}~{\rm cm}^{-2} {\rm meV}^{-1}$, which for the range
of densities considered in Ref.\cite{pudalov00} (i.e.
$n=0.7-1.3\times 10^{11}{\rm cm}^{-2}$) gives for the Fermi energy 
$\epsilon_F =5.1-9.5 {\rm K}$.
By assuming $g=2$ a magnetic field of $1~$T gives 
$\omega_s =0.6$K. For the spin-orbit
interaction, we take from Ref.\cite{pudalov00} $\alpha =6\times
10^{-6}$K cm, which
gives a spin-orbit energy $\alpha p_F =3.36$ K. 
Considering transport introduces another scale, 
which we characterize by the scattering rate
$1/\tau$.
We assume throughout this paper that $\epsilon_F \tau $ is larger than unity
as is appropriate for the metallic regime (in units with $\hbar =1$).
A realistic mobility of $4\times 10^4 {\rm cm}^2/$Vs 
gives then a relaxation rate $1/\tau \sim 1-2 $K.
In the following we choose parameters having the Si systems in 
mind\cite{gallium}.
Measuring energies in units of $\epsilon_F$ one has that $1/\tau =0.1-0.4$, 
$\alpha p_F = 0.3-0.7$ and $\omega_s = 0.06-0.15$ per Tesla.

A simple estimate for the conductivity and its anisotropy may be obtained 
with the semi-classical 
Boltzmann equation in the relaxation time approximation.
Assuming a single transport time, i.e.  
momentum- and band-independent,
the conductivity is
\begin{equation}
\sigma_{ij} = e^2 \tau_{\rm tr} \sum_{s = \pm} N_{s} \langle v^i_s v^j_s \rangle_{FS}
\end{equation}
where the sum is over the two bands, $N_{\pm}$ is the density of states,
${\bf v}_{\pm} = \nabla_{\bf p} E_{\pm}({\bf p})$ the velocity,
and $\langle \dots \rangle_{FS}$ the average over the Fermi surface.
It is apparent that the transport time drops, when we consider the conductivity relative to its
value in the absence of magnetic field and spin-orbit coupling,
$\sigma_0 = e^2 N_0 v_F^2 \tau_{\rm tr}$.
\begin{figure}
\noindent
\begin{minipage}[t]{0.98\linewidth}
{\centerline{\epsfxsize=7cm\epsfbox{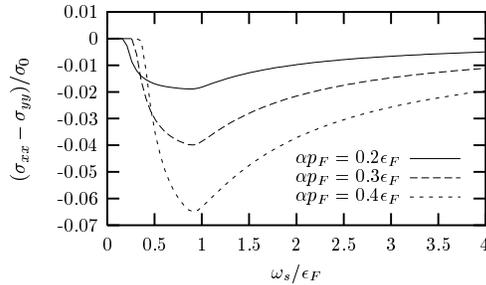}}}
\caption{ 
Anisotropy in the magneto-conductance as obtained from the
Boltzmann equation in the relaxation time approximation.
}
\label{amr-bo}
\end{minipage}
\end{figure}
In Fig.\ref{amr-bo} 
we plot the anisotropy in the magneto-conductance for various strengths of
the spin-orbit coupling. One observes a negative anisotropy, i.e. the 
conductivity
perpendicular to the field is larger than that parallel to the field, with a 
maximum
in the region $\omega_s \sim \epsilon_F$.
Some insight in the results reported in Fig.\ref{amr-bo}
is obtained from the analytic expressions in the weak ($\omega_s < \epsilon_F$)
and strong magnetic field limits.
For the given dispersion relation the velocities are
${\bf v}_{\pm} = {\bf p}/m \pm \alpha {\bf e}_{\bf p}$, with
${\bf e}_{\bf p} = (\alpha {\bf p} -\omega_s {\bf e}_y)/\Omega ({\bf p})$.
In the weak magnetic field limit, one finds that  only the anomalous part of
the
velocity contributes to the anisotropy in the conductance. We obtain then
\begin{equation}
\sigma_{xx} - \sigma_{yy} = 2 e^2 N_0 \tau_{\rm tr} \alpha^2
\langle ({\bf e}_{\bf p }\cdot {\bf e}_x )^2 -
( {\bf e}_{\bf p} \cdot {\bf e}_y )^2 \rangle_{FS}
.\end{equation}
For $\alpha p_F >\omega_s$ the polarization vector ${\bf e}_{\bf p}$ makes a 
full rotation when 
averaging over the Fermi surface. In the limit $\omega_s > \alpha p_F$ however 
the polarization
vector becomes locked with ${\bf e}_{\bf p} \approx -{\bf e}_y$. 
It is clear that 
 one then finds
a negative anisotropy in the conductance.
Going explicity through the algebra we arrive at
\begin{equation}
{ \sigma_{xx} - \sigma_{yy} \over \sigma_0 } = 
{1\over 2} \left( {\alpha p_F \over \epsilon_F } \right)^2
{ 1 - (\omega_s/  \alpha p_F )^2 
\over (\omega_s / \alpha p_F )^2 }
\Theta\left( {\omega_s \over \alpha p_F} -1  \right)
,\end{equation}
which has an edge at $\omega_s = \alpha p_F$ and
becomes constant when $\omega_s \gg \alpha p_F$.
In the strong field limit the arguments are different.
For large field only the lower band is occupied. 
The Fermi surface again is almost rotational symmetric, 
but there is a weak elliptic distortion
which leads to an anisotropy:
\begin{equation}
{ \sigma_{xx} - \sigma_{yy} \over \sigma_0 } =
-{1\over 2} {(\alpha p_F )^2 \over \omega_s \epsilon_F }
.\end{equation}
From these considerations we find that the anisotropy 
for $\omega_s \sim \epsilon_F$ scales as $(\alpha p_F/\epsilon_F)^2$.

In order to have a more microscopic calculation of the conductivity
we now go beyond the relaxation time approximation.
We introduce  a disorder potential $U$, which will be treated in the 
self-consistent Born approximation and
the conductivity will be calculated using the Green function formalism.
For simplicity we assume a impurity potential with short range 
Gaussian correlations
\begin{equation}
\langle U({\bf x} ) U({\bf x }')\rangle  = {1\over 2\pi N_0 \tau}
\delta( {\bf x} - {\bf x}' )
.\end{equation}
The self-energy is determined self-consistently by the equation
\begin{equation}
 \label{selfenergyequation}
\Sigma = {1\over 2\pi N_0 \tau} \sum_p G(p)
,\end{equation}
where $G(p)$ is the Green function. Notice that $\Sigma$ and $G(p)$
are $2 \times 2$ matrices.
Expanding the self-energy
in Pauli matrix components, it turns out that
the self-energy has no $\sigma_2$ and $\sigma_3$ components.
As a result the self-energy will have the form
$\Sigma = \Sigma_0 \sigma_0 +\Sigma_1 \sigma_1$.
The real part of the self energy $\Sigma_0$ shifts the energy spectrum
by a constant. Since we have to adjust the chemical potential in order to keep
the particle number at a given value $\Re \Sigma_0$ can  be safely neglected.
The real part of $\Sigma_1$ gives rise to a renormalization of the Zeeman energy.
For the imaginary part, we find 
in the limit of weak disorder
\begin{eqnarray}
\Im \Sigma^R_0  = -1/2\tau_0 &\approx & -(N_+ +N_-)/4 N_0 \tau \\
\Im \Sigma^R_1  = -1/2\tau_1 &\approx & -(N_- -N_+)/4 N_0 \tau
\end{eqnarray}
The sum or difference 
$1/\tau_{\mp}= 1/\tau_0 \pm 1/\tau_1$ are roughly the scattering rates for
the two subbands.
For weak magnetic field $(\omega_s < \epsilon_F)$, the density of states in the
two subbands are identical and therefore
$ 1/\tau_\pm = 1/\tau_0 = 1/\tau$.
In the strong field limit $(\omega_s > \epsilon_F)$
the upper band is depopulated, so that
$ 1/\tau_1 = 1/\tau_0 = 1/2 \tau$.

The current operator, which is necessary for the evaluation of the
conductivity, is modified in the presence of the spin-orbit interaction:
\begin{equation}
\label{3}
 {\bf j} =\frac{ {\bf p}}{m}-\alpha{\mathbf \sigma}\wedge {\bf e}_z.
\end{equation}
The conductivity is obtained by the formula 
\begin{eqnarray}
\sigma_{ij}&=&\frac{e^2}{4\pi}\sum_{{\bf p}}
{\rm Tr}\left[ j^iG^RJ_{RA}^jG^A-
j^iG^RJ_{RR}^jG^R \right. \cr
&+&\left. j^iG^RJ_{RA}^jG^A-
j^iG^AJ_{AA}^jG^A\right].
\end{eqnarray}
The dressed vertex $J^j$ depends upon whether it is connected to a pair of
retarded and advanced Green functions or 
a pair of Green functions with equal analytic properties.

In order to evaluate the conductivity, we have to determine the
renormalized vertices and perform the momentum integrals. 
Before showing the complete results it is
useful to  consider the conductivity
in the absence of vertex corrections and compare with the above Boltzmann
equation analysis. 
\begin{figure}
\noindent
\begin{minipage}[t]{0.98\linewidth}
{\centerline{\epsfxsize=7cm\epsfbox{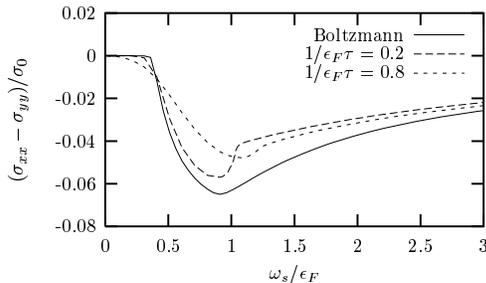}}}
\caption{ 
Anisotropy in the magneto-conductance for $\alpha p_F = 0.4 \epsilon_F$, 
$1/(\epsilon_F \tau)= 0.4$;
the full line is the result from the Boltzmann equation. 
The dashed lines result from the
Kubo formula, when vertex corrections are neglected,
but using the 
self-consistently
determined self-energies in the Green functions.
}
\label{novertex}
\end{minipage}
\end{figure}
The result is shown in Fig.(\ref{novertex})(dashed lines). 
The conductivity
may be expressed as a sum of intra- and interband contributions.
Because of the
gap between the two spin subbands, the interband terms do not contribute much 
unless
the disorder is strong enough to produce a broadening of the Fermi surface
of the two bands. The condition for weak disorder is then $\Omega \tau \gg 1$.
In the weak disorder limit one reproduces -- when $\omega_s <\epsilon_F$ -- the results
obtained from the Boltzmann equation (the solid line).
For large magnetic field the Boltzmann result is {\sl not} reproduced, 
even for
very weak disorder. The reason is that even an isotropic scattering potential
generates
an anisotropic quasi-particle lifetime as demonstrated here below.
Let us denote the scattering probability from state ${\bf p}$ to
${\bf p}'$ by $W_{{\bf p}{\bf p}'}$.
When going from the spin- to the eigenstate basis 
the scattering probability in the lower band
becomes
\begin{equation} \label{scattering}
W_{{\bf p}{\bf p}'} \to 
W_{{\bf p}{\bf p}'} {1\over 2}\left( 1 + 
{\bf e}_{\bf p} \cdot {\bf e}_{{\bf  p}'}  \right)
\approx 
W_{{\bf p}{\bf p}'}
\left(1 -
\alpha^2 { (p_x - p_x')^2 \over 4 \omega_s^2}\right)
.\end{equation}
From this anisotropic scattering probability the inverse lifetime 
of an electron in state ${\bf p}$
is determined as
$1/\tau_{\bf p} \approx [1-(\alpha p_x)^2/4 \omega_s^2]/\tau$
which definitely depends on the position on the Fermi surface.
The anisotropic scattering rate then also contributes to the anisotropy 
in the magneto-conductance.

\begin{figure}
\noindent
\begin{minipage}[t]{0.98\linewidth}
{\centerline{\epsfxsize=5.5cm\epsfbox{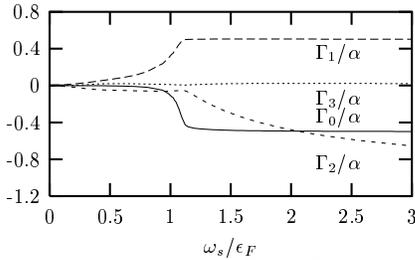}}}
\caption{ 
The dressed vertices $\Gamma_0 $ \dots $\Gamma_3$  in units of $\alpha$ as 
a function of the
magnetic field. Here we chose $\alpha p_F = 0.4 \epsilon_F$ and 
$1/(\epsilon_F \tau) = 0.4$.
Remember that the bare vertices are
$\gamma_0 = \gamma_3 = 0$ and $\gamma_1 = -\gamma_2 = \alpha$;
the asymptotic values in the strong field limit are
$\Gamma_1 = - \Gamma_0 =\alpha/2$, $\Gamma_2 = -\alpha$, and $\Gamma_3 = 0$. 
Notice that $\Gamma_2$ reaches the asymptotic value for high magnetic 
field only very slowly.
}
\label{figr1}
\end{minipage}
\end{figure}

\begin{figure}
\noindent
\begin{minipage}[t]{0.98\linewidth}
{\centerline{\epsfxsize=7cm\epsfbox{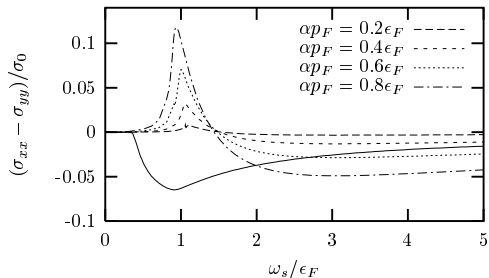}}}
\caption{Anisotropy in the magneto-conductance 
for $1/(\epsilon_F \tau) = 0.2$ and various spin-orbit energies.
For comparison we included the results of the Boltzmann theory 
($\alpha p_F = 0.4 \epsilon_F$;
full line).  
For not too strong fields, the vertex corrections change the 
sign of the effect.}
\label{withvertex}
\end{minipage}
\end{figure}

We are now ready to consider the vertex corrections. For the choice of
disorder potential we have made, these corrections
usually vanish expressing the fact that the relaxation of momentum
is governed by the quasiparticle lifetime. However, in the present problem
there are specific current vertices due to the spin-orbit interaction
which acquire a vertex correction. The vertex corrections are  obtained
in the standard way by solving the equation
\begin{equation}J_{RA}^i= j^i + {1\over 2\pi N_0 \tau} \sum_{\bf p}  G^R J_{RA}^i
 G^A.
\end{equation}
By separating  the  $p$-dependent (i.e., proportional to
the ${\bf p}$ vector) and  $p$-independent parts
$j^i  =  p^i/m + \gamma^i$ and $J^i = J_0^i(p)+ \Gamma^i$,
one notices that as a consequence of isotropic scattering, the
$p$-dependent part is not dressed, i.e., $J^i =p^i/m + \Gamma^i$. 
The equation for the momentum independent part of the current vertex reads
\begin{equation}
\Gamma^i_{s s'}=\tilde\gamma_{s s'}^{i}+
\frac{1}{2\pi N_0 \tau }\sum_{\bf p} \sum_{a b}G^R_{sa}\Gamma_{a b}^i 
G^A_{b s'}.
\end{equation}
We included here the matrix (=spin) indices $s$, $s'$, $a$, and  $b$.
The quantities $\tilde \gamma^{i}$ are a sum of the bare vertices $\gamma^i$ 
and a term which is generated by ${\bf p}/m$,
\begin{equation}
\tilde\gamma^{i}_{ss'}= \gamma^i_{ss'} +\frac{1}{2\pi N_0\tau} \sum_{\bf p} 
\sum_a  G^R_{s a} (p^i/m ) G^A_{a s'}.
\end{equation} 
The above equation may be solved by expanding all matrices in terms of the
Pauli matrix basis, e.g., $\Gamma^i=\sum_{\mu =0,3}\sigma_{\mu}\Gamma^i_{\mu}$.
By skipping the details of the derivation we present the numerical results
for the renormalized vertices and the conductivity in Figs.(\ref{figr1}) and
(\ref{withvertex}), respectively.
One observes that for weak magnetic field the quantities $\Gamma$
become very small so that ${\bf J}_{RA} \approx {\bf p}/m$ i.e. 
the anomalous velocity is cancelled by the vertex corrections!
Only for fields which are at least of the order of the Fermi energy
does an anomalous velocity contribution survive.

Concerning the conductivity the most striking result is that  
the vertex corrections,  for not too strong magnetic field,
change the sign of the anisotropy.
(See Fig.\ref{withvertex}).
To understand the effect of the vertex correction, it is again useful
to consider the scattering rate in the eigenstate basis,
see eq.(\ref{scattering}). One then
observes  that along the ${\bf x}$ direction, {\sl forward} scattering is 
enhanced compared to {\sl backward} scattering, while along the
${\bf y}$ direction there is no difference
between {\sl forward} and {\sl backward}. 
Because vertex corrections  
 increase conductivity if {\sl forward} scattering is favoured,
one expects that vertex corrections enhance 
$\sigma_{xx}$ most contributing  to $\Delta\sigma$ with a positive term, which
for low and moderate magnetic  fields overcomes the negative
contribution discussed previously.
 
We now comment on our results in the light of the available experiments.
Both in the relaxation time approximation and in the Green function calculation
the maximum anisotropy in the conductivity
occurs at $\omega_s \sim \epsilon_F$ irrespective of the value
of the spin-orbit energy $\alpha$. 
This means that
the energy of the peak scales with the electron density. This is in agreement
with the experiments\cite{pudalov00}. 
A second feature is that the peak strength is controlled
by the spin-orbit energy in units of $\epsilon_F$. Hence the peak strength
must  scale with the inverse density, so that the effect is expected to 
decrease going
more deeply into the metallic regime. This again is in agreement with the
experimental findings\cite{pudalov00,khrapai00}.
Concerning the sign of the effect, inclusion of the vertex corrections
in the Green function calculation spoils the 
agreement with the experiments. A possible reason for this
discrepancy may be due to our simplifying assumption  of a short range 
p-independent disorder potential. Our analysis has shown in fact that
the sign of the anisotropy depends crucially on the {\sl effective}
p-dependence of the scattering rate in the eigenstate basis.
It is then reasonable to expect that the specific choice
of the scattering potential may indeed play a role in determining
the sign of the effect a low fields. This analysis, while it is worth
pursuing, is however beyond the scope of the present paper.
The anisotropy in the magnetoresistance observed
in $GaAs$ heterostructures\cite{papadakis99}
presents a rather complex pattern. Two crystallographic
directions with different mobilities have been investigated. The one with
higher mobility, 
shows anisotropy with both $\Delta \sigma <0$ at not too large magnetic
fields and $\Delta \sigma >0$ at larger fields.
At low fields (where $\Delta \sigma < 0$) the value of the field at which the
anisotropy is maximum seems to scale with the density
(cf. right column of Fig.2 of Ref.\cite{papadakis99}).
At this stage one can only speculate that the actual sign of the
anisotropy may depend on the interplay of the
details  of the scattering potential with the spin-orbit coupling effect
discussed here. 

In conclusion, our theory explains how
the spin-orbit coupling may give rise to  
the anisotropy in the conductivity, although the interpretation
of the experiments may require more realistic 
models of disorder, as for example that due to
scattering by impurities located outside the 2D plane, which  
favours small-angle scattering.
It is also our hope that this work will stimulate more experimental
effort toward the investigation of anisotropic conductivity as a function
of the scattering potential or in sample with different mobilities.

This work was partially supported by MURST under contract no. 9702265437  
(R.R. and C.C.), by INFM under the PRA-project QTMD (R.R.) and the European  
Union TMR program (M.L.) and the DFG through SFB 484 (P.S. and R.R.).

\end{multicols} 

\begin{references}
\bibitem{gallium}For hole carrier $GaAs$ heterostructures one has 
                a density-of-states of $N_0 = 1.8\times10^{11}~{\rm cm}^{-2} 
                {\rm meV}^{-1}$ with $m=0.38 m_0$
                and for the density range $n=1-6 \times 10^{10}{\rm  cm}^{-2}$ 
                considered in the
                experiments \cite{papadakis99} one has a Fermi energy 
                $\epsilon_F =0.6-3.8$ K.
\end{references}
\end{document}